\documentclass[aps,prd,showpacs,twocolumn,superscriptaddress,nofootinbib,preprintnumbers]{revtex4-1} 

\makeatletter
\def\p@subsection{}
\makeatother

\usepackage{color,graphicx,amsmath,amssymb,lipsum}
\usepackage[colorlinks=true,citecolor=blue,linkcolor=blue,urlcolor=blue, backref=false,pdfborder={0 0 0}]{hyperref}

\usepackage[normalem]{ulem}
\usepackage[utf8]{inputenc} 
\usepackage{mathtools}

\usepackage{float}

\def\mnras{\rm{MNRAS}}

\def\apj{\rm{ApJ}}                 
\def\jcap{\rm{JCAP}}

\usepackage{ulem}
\usepackage{diagbox}

\usepackage[dvipsnames]{xcolor}
\usepackage{mathrsfs}

\newcommand{\be}{\begin{equation}}
\newcommand{\ee}{\end{equation}}
\newcommand{\beqa}{\begin{eqnarray}}
\newcommand{\eeqa}{\end{eqnarray}}

\newcommand{\bseq}{\begin{subequations}}
\newcommand{\eseq}{\end{subequations}}

\renewcommand{\ln}{\mathop{\rm ln}\nolimits}

\newcommand{\edit}[1]{{#1}}

\def\gsim{\raise0.3ex\hbox{$\;>$\kern-0.75em\raise-1.1ex\hbox{$\sim\;$}}}
\def\lsim{\raise0.3ex\hbox{$\;<$\kern-0.75em\raise-1.1ex\hbox{$\sim\;$}}}

\def\beqn#1{\begin{equation}\label{#1}}
\def\eeqn{\end{equation}}

\def\beqa#1{\begin{eqnarray}\label{#1}}
\def\eeqa{\end{eqnarray}}

\def\kmax{{k_\text{max}}}
\def\hMpc{h{\text{Mpc}}^{-1}}
\def\Mpch{h^{-1}{\text{Mpc}}}

\def\Z2{$\mathcal{Z_2}$}

\newcommand {\ignore}[1]{}

\begin{document}

\preprint{MIT-CTP/5731}

\title{{\Large Suppression without Thawing:}\\ Constraining Structure Formation and Dark Energy with Galaxy Clustering
}

\author{Shi-Fan Chen}
\email{sfschen@ias.edu}
\affiliation{School of Natural Sciences, Institute for Advanced Study, 1 Einstein Drive, Princeton, NJ 08540, USA}

\author{Mikhail M. Ivanov}
\email{ivanov99@mit.edu}
\affiliation{Center for Theoretical Physics, Massachusetts Institute of Technology, 
Cambridge, MA 02139, USA}

\author{Oliver H.\,E. Philcox}
\email{ohep2@cantab.ac.uk}
\affiliation{Simons Society of Fellows, Simons Foundation, New York, NY 10010, USA}
\affiliation{Center for Theoretical Physics, Columbia University, New York, NY 10027, USA}
\author{Lukas Wenzl}
\email{ljw232@cornell.edu}
\affiliation{Department of Astronomy, Cornell University, Ithaca, NY, 14853, USA}

\begin{abstract}
\noindent We present a new perturbative full-shape analysis of BOSS galaxy clustering data, including the full combination of the galaxy power spectrum 
and bispectrum multipoles, baryon acoustic oscillations, and cross-correlations with the gravitational lensing of cosmic microwave background measured from \textit{Planck}. Assuming the $\Lambda$CDM model, we constrain the matter density fraction $\Omega_{\rm m} = 0.3138\pm 0.0086$, the Hubble constant $H_0=68.23\pm 0.78\,\mathrm{km}\,\mathrm{s}^{-1}\mathrm{Mpc}^{-1}$, 
and the mass fluctuation amplitude $\sigma_8=0.688\pm 0.026$
(equivalent to $S_8 = 0.703\pm 0.029$). 
Cosmic structure at low redshifts appears 
suppressed with respect to the \textit{Planck} $\Lambda$CDM concordance model at $4.5\sigma$. We explore \edit{the implications of this data set for} the recent DESI BAO preference for dynamical dark energy (DDE): the BOSS data combine with DESI BAO and PantheonPlus supernovae competitively compared to the CMB,
yielding no preference for DDE, but the same $\sim 10\%$ suppression of structure, with dark energy being consistent with a cosmological constant at 68\% CL. Our results suggest that either the data contains residual systematics, or more model-building efforts may be required to restore cosmological concordance. 
\end{abstract}

\maketitle

\textit{Introduction. ---} Observational and theoretical efforts over the last three decades have led to the establishment of the standard model of cosmology: $\Lambda$CDM. This model successfully fits a wide range of cosmological data, in particular the various correlators of cosmological fluctuations traced by the cosmic 
microwave background (CMB) anisotropies and the large-scale structure of the Universe~\citep[e.g.,][]{Planck:2018vyg,eBOSS:2020yzd}.

Despite its phenomenological successes, the $\Lambda$CDM model suffers from significant theoretical questions. Many of its core ingredients, such as cosmic inflation, dark matter, and dark energy are, at best, highly exotic. The latter is particularly puzzling from the theoretical viewpoint. 
The simplest explanation for dark energy is the cosmological
constant, which gives rise to the naturalness paradox that shatters 
the fundamental pillars of physics: symmetry-based selection rules 
and dimensional analysis~\cite{Weinberg:1988cp,Burgess:2013ara}. Whilst anthropic~\cite{Weinberg:1987dv} and landscape~\cite{Polchinski:2006gy} explanations are possible, the cosmological constant problem still poses a formidable conceptual challenge in fundamental physics. 
This challenge is particularly relevant given the possible ($\gtrsim 2.5 \sigma$) evidence for dynamical dark energy (DDE, or $w_0w_a$CDM) recently reported by the Dark Energy Survey Instrument (DESI)
collaboration~\cite{DESI_BAOGAL,DESI_BAOLyA,DESI:2024mwx}. 

\begin{figure*}
    \centering
    \includegraphics[width=0.8\textwidth]{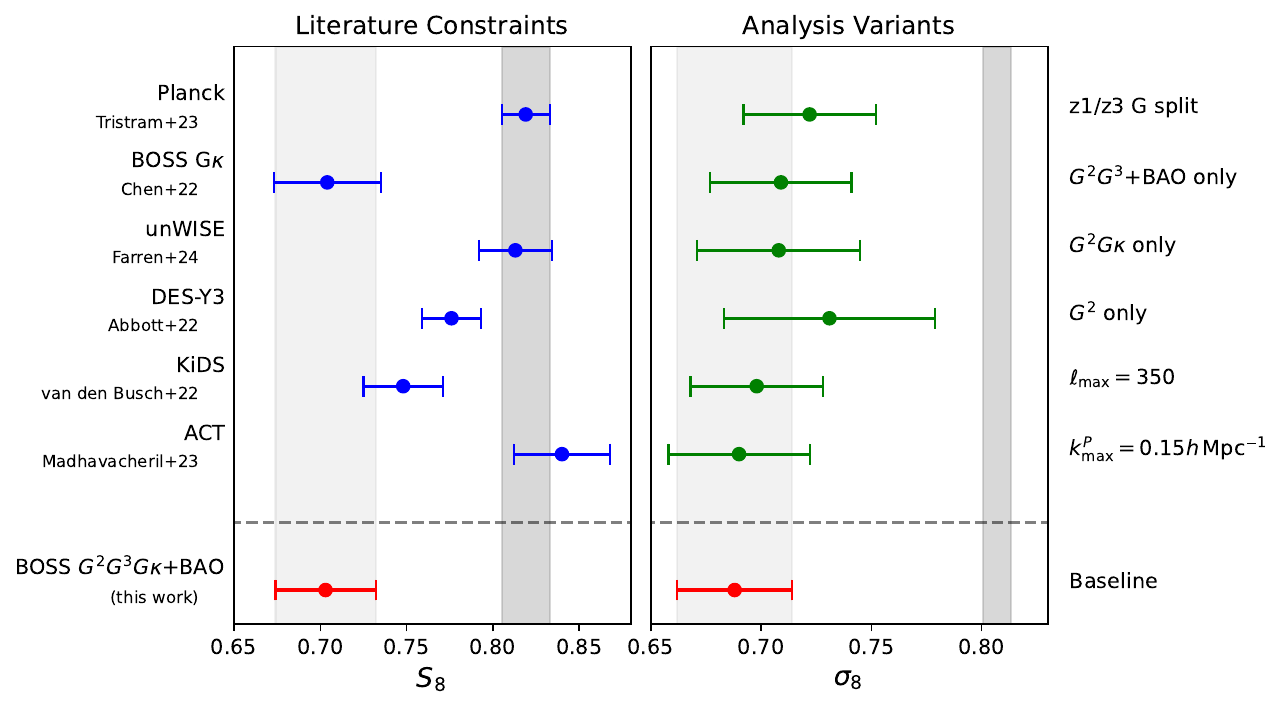}
    \caption{\textbf{Left}: A comparison on various $S_8$ results available in the literature \citep{Chen22,ACT:2023oei,DES:2021wwk,Busch:2022pcx,ACT:2023kun,Tristram:2023haj}, with our measurement (bottom) including galaxy two- and three-point information ($G^2$ and $G^3$), cross-correlation with lensing ($G\kappa$), and BAO. \textbf{Right}: Dependence of our results on analysis choices including choice of galaxy split, dataset, maximum cross-correlation scale, \edit{and maximum scale for the power spectrum multipoles (also dropping $Q_0$)}.}
    \label{fig: s8_plot}
\end{figure*}

In addition to DDE, the data contain other anomalies whose presence could signal the breakdown of cosmological concordance. The most prominent
is the Hubble tension, \textit{i.e.}\ the apparent disagreement between the direct and indirect measurements of the Hubble constant $H_0$, a proxy for the age of the Universe~\cite{SH0ES}. 
Another important 
anomaly is the disagreement of 
direct and indirect probes of the growth of structure encoded by the mass fluctuation amplitude $\sigma_8$, or the related structure growth parameters $S_8\equiv(\Omega_{\rm m}/0.3)^{1/2}\sigma_8$ and $f\sigma_8$ (where $f$ is the redshift-dependent logarithmic growth factor)~\cite{DiValentino:2020vvd,Nguyen:2023fip}. This discrepancy is observed in multiple independent low-redshift datasets \cite{Abdalla:2022yfr} (Fig.\,\ref{fig: s8_plot}): cluster counts \citep[e.g.,][]{Planck:2015lwi,Bolliet:2019zuz}, weak lensing measurements~\citep[e.g.,][]{KiDS:2020suj,DES:2021wwk}, CMB lensing cross-correlations \citep[e.g.,][]{ACT:2023ipp,White:2021yvw}, and galaxy clustering in redshift space~\citep[e.g.,][]{Nguyen:2023fip,Ivanov23,Philcox22,Chen22,Chen:2021wdi,Ivanov:2021zmi}, though there exist some outliers \citep{BOSS:2016psr,Horowitz:2016dwk,Kobayashi:2021oud,Dalal23,Miyatake23,Yu:2022tzw,DAmico:2022osl,ACT:2023oei,ACT:2023skz}. In general relativity the expansion history and growth of structure are intricately related through the equations of motion, and the accumulation of cosmological tensions raise a natural question: do they all point to a particular new physics model in a correlated fashion? 
This \textit{Letter} addresses 
this question focusing on the case of DDE and the $\sigma_8$ tension. 
\edit{This question is natural as the best-fit 
DESI DDE model implies a nontrivial 
evolution of the DDE equation of state $w(z)$, which suggests a suppression of clustering at $z>0.5$
and enhancement for $z<0.5$: it is interesting to see if we can test the DESI findings using full-shape information beyond the BAO in galaxy clustering and whether the $\sigma_8$ discrepancy remains in a more flexible cosmological background which appears to be preferred by recent data.}

We present an independent reanalysis of the galaxy clustering data from the Baryon Acoustic Oscillation Spectroscopic Survey (BOSS) in combination with \textit{Planck} CMB lensing, in an attempt to link the possible $\sigma_8$ tension found in these data with the hints of DDE reported by DESI. Using methodologies developed in previous works \citep{Philcox22,Ivanov23,Ivanov:2019pdj,Ivanov:2019hqk,Ivanov:2021fbu,Philcox20}, we measure the BOSS three-dimensional redshift-space power spectrum and bispectrum multipoles, and post-reconstructed BAO data. We also include the angular cross correlation between 
the BOSS galaxies and \textit{Planck} CMB lensing following \cite{Wenzl24} (see also \cite{Pullen16,Singh17,Doux18,Singh19,Darwish21,Chen22,Wenzl24,Wenzl24b}). 
For the first time, we consistently analyze all of these observables within the effective field theory (EFT)-based full-shape (FS) framework~\cite{Ivanov:2019pdj,DAmico:2019fhj,Chen:2021wdi}. 
Our first important result is that, when combined with a BBN prior on the baryon density, this dataset yields a measurement of matter clustering amplitude 
$\sigma_8$ discrepant with the \textit{Planck} concordance value at the $4.5\sigma$ level (cf.\,Fig.\,\ref{fig: s8_plot}). This represents the strongest evidence for the $\sigma_8$ tension from the BOSS dataset to date. 

In the second part of this \textit{Letter}, \edit{we open up the parameter space to constrain DDE and explore whether the $\sigma_8$ tension can be explained by the DDE model suggested by the combination DESI$+$Supernovae (SNe) and CMB.} Specifically, we analyze 
the BOSS and lensing data described above, the DESI BAO data at redshift $z>0.8$, and PantheonPlus SNe data assuming the DDE model. \edit{Our dataset constrains DDE competitively compared to DESI BAO but does not display any evidence
for DDE.} We find that DDE does not restore concordance between 
galaxy clustering data and the primary CMB. The optimal values of $\sigma_8$ in our DDE analysis is still in $4.5\sigma$ tension with \textit{Planck}, though the $H_0$ is consistent with \textit{Planck} and not with the value implied by the Cepheid-calibrated distance 
ladder~\cite{SH0ES} (though in better agreement with \citep{Freedman:2022uxj}). The combination of the above results suggests that internal tensions between the datasets seem to pull cosmological parameters in directions uncorrelated with each other; this could motivate more efforts from the 
model building perspective, as well as searches for systematic
effects in the full combination of the large-scale structure 
data. 

\vskip 8pt
\textit{Data. ---} Our primary dataset is the clustering of galaxies from the twelfth data release of the BOSS survey \cite{Reid16,BOSS:2012dmf}. These galaxies are observed in both the northern (NGC) and southern (SGC) galactic caps and are composed of the \texttt{LOWZ} and \texttt{CMASS} samples, each of which are restricted to the redshift ranges $0.15 < z < 0.43$ and $0.43 < z < 0.70$ in order to avoid overlap.
Combining both galactic caps, the \texttt{LOWZ} and \texttt{CMASS} catalogs cover $8,579$ and $9,493$ deg$^2$ with $361,762$ and $777,202$ galaxies, respectively. The complete DR12 catalogs also contain galaxies in two chunks \texttt{LOWZE2} and \texttt{LOWZE3} selected using different criteria than the main \texttt{LOWZ} sample. These are often combined with the main samples in order to maximize the survey volume, but, since the selections imply quite different galaxy properties, we will instead omit them in this work (this choice was made also in pre-DR16 BOSS analyses \citep[e.g.,][]{GilMarin16}). \edit{In addition, the systematics weights for the BOSS galaxies were computed for \texttt{LOWZ}/\texttt{CMASS}, leading to potential biases due to redshift-dependence when using them for the \textbf{z1}/\textbf{z3} samples \cite{Chen22}.}

To characterize the clustering of the above galaxy samples, we utilize the power spectrum and bispectrum statistics, measured using the window-free estimators derived in \citep{Philcox21a,Philcox21b,Ivanov23} (now implemented in the \textsc{PolyBin3D} code \citep{Philcox:2024rqr}). We include the standard systematic and \edit{Feldman-Kaiser-Peacock (FKP)} weights constructed by BOSS~\cite{Reid16}, which imply that the power spectrum probes clustering at an effective redshift $z_{\rm eff} = \int {\rm d}V\,z\,\bar{n}^2(z) / \int {\rm d}V\bar{n}^2(z)$, equal to $0.316$ ($0.555$) for the \texttt{LOWZ} (\texttt{CMASS}) sample, where $\bar{n}$ is the weighted galaxy number density. The same weights applied to the bispectrum would result in a different effective redshift (instead weighted by $\bar{n}^3$); to ameliorate this, we an additional redshift weight $w_B(z) \equiv \bar{n}(z)^{-1/3}$ when computing the bispectrum. We additionally include the real-space power spectrum proxy $Q_0$ (roughly the power spectrum perpendicular to the line of sight) estimated from the redshift space multipoles~\cite{Ivanov:2021fbu}.

In addition to the power spectrum and bispectrum, we also include post-reconstruction BAO measurements from the BOSS galaxies \edit{(which fold in additional information from higher-order statistics \citep{Schmittfull:2015mja})}. Since this signal does not depend strongly on galaxy properties, we will use measurements of the BAO scale from the \textit{combined} BOSS sample covering the full survey area and redshift range, including the \texttt{LOWZE2} and \texttt{LOWZE3} samples omitted in the above full-shape analysis (specifically, those from \citep{Philcox20}). This combined sample is split into non-overlapping redshift bins $0.2 <z < 0.5$ (\textbf{z1}) and $0.5 <z < 0.75$ (\textbf{z3}) chunks following~\cite{BOSS:2016wmc,BOSS:2016psr} with effective redshifts of $z_{\rm eff}= 0.38$ and $0.59$, respectively. We compute the total covariance of the above statistics \edit{(both spectra and BAO parameters)} using measurements from the 2048 public MultiDark Patchy  mocks~\cite{Patchy,Rodriguez-Torres:2015vqa}.

\begin{figure}
\includegraphics[width=0.49\textwidth]{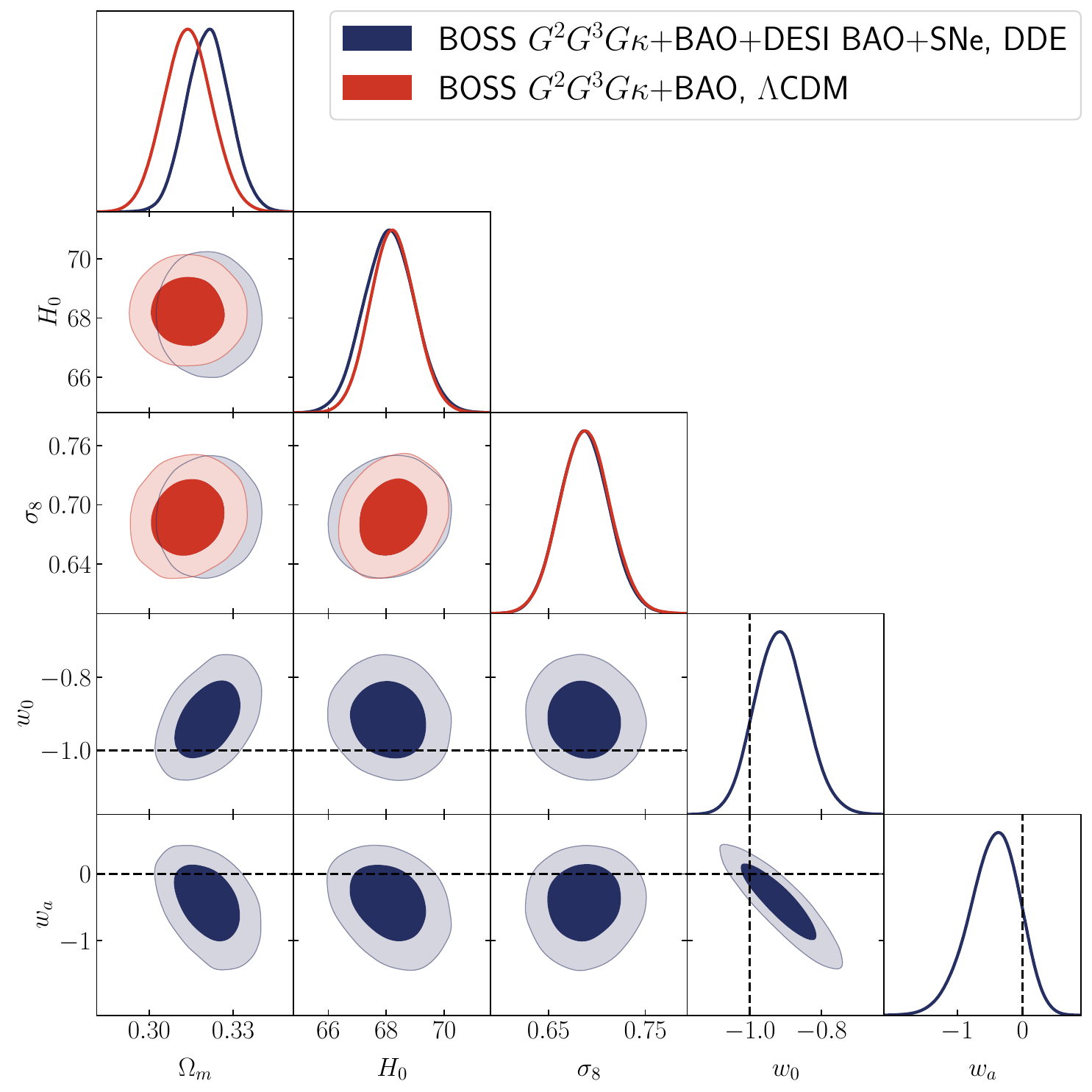}
   \caption{Constraints on cosmological 
   parameters from our baseline dataset (BOSS two- and three-point galaxy clustering correlations, cross correlations of galaxies and CMB lensing and BAO; BOSS $G^2G^3G\kappa$+BAO) on the standard $\Lambda$CDM model (red), as well as those on the dynamical dark energy (DDE) model from the baseline data in addition to DESI BAO (at $z>0.8$) and PantheonPlus supernovae (blue). Dashed lines mark the $\Lambda$CDM values of the dark energy equation of state parameters $w_0$ and $w_a$, corresponding to the cosmological constant. 
    } \label{fig:main}
\end{figure}

In order to measure the lensing cross correlation with CMASS and LOWZ galaxies we use publicly-available CMB lensing maps reconstructed from \textit{Planck} data. Specifically, we use the PR4 map introduced in \cite{Carron22}, which uses the updated \texttt{NPIPE} pipeline and slightly more data than previous releases, leading to a $20\%$ improvement in signal-to-noise compared to PR3. We compute the galaxy cross-correlations $C^{\kappa g}_\ell$ with the lensing convergence $\kappa$ and its bandpower window (analytic) covariance using \texttt{NaMaster} \cite{Alonso19,GarciaGarcia19,Nicola21} adopting the same numerical pipeline described in \cite{Wenzl24}---who also conducted extensive systematics tests on these data---except that we split the cross correlations by galactic cap. We treat the covariance independently from that derived from the three-dimensional clustering of galaxies since the mode overlap is negligible \cite{Taylor22}. We re-weight the galaxies when computing $C^{\kappa g}_\ell$ by the ratio of the galaxy and CMB-lensing kernels to homogenize the effective redshifts probed with the galaxy power spectrum \cite{Chen22}.

Finally, in addition to BOSS-volume data, we will supplement our analysis with constraints on cosmological expansion at lower and higher redshfts obtained from Supernovae Type Ia (SNIa) and external BAO data. For the former we adopt the PantheonPlus dataset \cite{Pantheon+, SH0ES} (which constrains the redshift-dependence of luminosity distances) and BAO likelihoods \edit{from lower redshifts} for the 6-degree Field Galaxy Redshift Survey (6dFGS) \cite{6dFGS} and the Main Galaxy Sample (MGS) in SDSS DR7 \cite{MGS}, as implemented in \texttt{MontePython}~\cite{Audren:2012wb,Brinckmann:2018cvx}. For the latter, we use all galaxy BAO measurements from DESI with $z > 0.8$ \cite{DESI_BAOGAL} as well as the DESI Ly$\alpha$ measurement \cite{DESI_BAOLyA}. We also adopt the BBN baryon density constraint from the primordial deuterium abundance $\Omega_{\rm b} h^2 = 0.02268 \pm 0.00038$~\cite{Cooke:2017cwo,Ivanov:2019pdj}, and fix\footnote{See the Supplementary Materials for further details on this choice.} the spectral tilt to the \textit{Planck} best-fit value $n_{\rm s}=0.9649$. 

\vskip 8pt
\textit{Theoretical model. ---}
We begin with a short overview of the theory model used for galaxy clustering. At the background level, we adopt either a baseline $\Lambda$CDM model, or the popular \edit{DDE} extension parameterized by the equation of state $w(a) = w_0 + w_a (1 - a)$~\cite{Fang:2008sn}. 
We assume flat priors $w_0\in [-3, 1]$, 
and $w_a \in  [-5,5]$ \edit{(which do not impact the analysis), and use infinite flat priors for the remaining cosmological parameters}. 
We compute predictions for the redshift-space power spectrum and bispectrum, as well as the real-space matter-galaxy cross power spectrum, using the effective field theory of large scale structure (EFT)~\cite{Baumann:2010tm,Carrasco:2012cv,Ivanov:2022mrd} as implemented in the \texttt{CLASS-PT} code~\cite{Chudaykin:2020aoj} and associated public likelihoods which we slightly modify.\footnote{\url{https://github.com/oliverphilcox/full_shape_likelihoods}} Our modeling follows the the conventions of \cite{Chudaykin:2020aoj,Chudaykin:2020hbf,Philcox22} to which we refer the reader for further details. Key technical details
are given in the Supplemental Materials, which include references also to Refs.\cite{Blas:2015qsi,Blas:2016sfa,Vlah16,Ivanov:2018gjr,Chen24,Senatore:2014via,Baldauf:2015xfa,Vlah15,Scoccimarro:2015bla,Scoccimarro:1999ed,Ivanov23,Limber53,LoVerde08,Mead:2015yca,Ivanov:2019pdj,Philcox22,Ivanov23,Beyond-2pt:2024mqz,Jackson:1971sky,Chudaykin:2020aoj}. 

Our baseline analysis of BOSS (which we dub $G^2G^3G\kappa$+BAO) uses redshift-space scale cuts $k_{\rm max}^{P_\ell}=0.2\,\hMpc$ and
$k_{\rm max}^{B_\ell}=0.08\,\hMpc$, and real-space scale cuts for $Q_0$ and $P_{mg}$ at  $k_{\rm max}^{\rm real}=0.4\,\hMpc$ (corresponding to angular cuts
$\ell_{\rm max}^{\texttt{LOWZ},\texttt{CMASS}}=400, 600$), all of which were validated \edit{against N-body simulation-based mocks in} in~\cite{Ivanov:2019pdj,Nishimichi:2020tvu,Chudaykin:2020hbf,Schmittfull:2020trd,Ivanov:2021kcd,Ivanov:2021fbu,Chen21,Philcox22,Chen:2021wdi,Chen22,Ivanov23,Beyond-2pt:2024mqz}.

\vskip 8pt
\textit{Results. ---}
We start by analyzing the BOSS FS and BAO data within $\Lambda$CDM. Our results are displayed in Fig.\,\ref{fig:main} and Tab.\,\ref{tab:cosmo_params}. We find that the optimal values of cosmological parameters are consistent with the \textit{Planck} baseline CMB values~\cite{Planck:2018vyg}, except for $\sigma_8$, which shows a $4.5\sigma$ disagreement. The tension with the ACT CMB lensing results~\cite{ACT:2023kun} has a similar strength: 4.3$\sigma$. For the $S_8$ parameter, the discrepancy is somewhat weaker, $3.8\sigma$, though our results appear in better agreement ($\lesssim 2.5\sigma$) with weak lensing measurements 
by DES~\cite{DES:2021wwk}, KiDS~\cite{KiDS:2020suj}, and HSC~\cite{Dalal23,Miyatake23} (Fig.~\ref{fig: s8_plot}).

Notably, the tension remains for different subsets of the underlying data (Fig.\,\ref{fig: s8_plot}). 
Our tests include: (a) adopting a more conservative choice of $\ell_{\rm max}=350$ for the lensing data (which yields $\sigma_8=0.698_{-0.03}^{+0.03}$); (b) fitting only the galaxy power spectrum multipoles using the Lagrangian EFT with the \texttt{velocileptors}
code\footnote{\url{https://github.com/sfschen/velocileptors/}} \cite{Chen20,Chen21,Chen:2021wdi,Maus24b} ($\sigma_8 = 0.731^{+0.048}_{-0.048}$); (c) fitting only the multipoles plus 
CMB-lensing cross correlation ($\sigma_8 = 0.708^{+0.037}_{-0.037}$) as in ref.~\cite{Chen22};
(d)
removing the galaxy-CMB lensing cross correlation ($\sigma_8 = 0.709_{-0.033}^{+0.031}$); (e) analyzing the \texttt{z1/z3} split 
of the BOSS data including \texttt{LOWZE2/LOWZE3} samples omitted in our main FS analysis ($\sigma_8=0.722_{-0.03}^{+0.03}$),
\edit{(f) using $\kmax=0.15~\hMpc$
for the power spectrum and excluding $Q_0$
($\sigma_8=0.690_{-0.033}^{+0.031}$).}
We find consistent results in all cases (Fig.~\ref{fig: s8_plot}). \edit{In particular, 
our results hold even if we use conservative scale cuts. This is because the bulk of the 
tension stems from large scales, $k<0.1~\hMpc$ where the effect of nonlinearities is smallest \cite{Chen:2021wdi,Philcox22}.} 

\begin{table}[ht]
    \centering
    \begin{tabular}{lclcl}
        \hline
        Parameter & \multicolumn{2}{c}{$\Lambda$CDM} & \multicolumn{2}{c}{DDE} \\
        \hline
        $\Omega_{\rm m}$ & $0.3138~(0.3188)_{-0.0089}^{+0.0087}$ && $0.3211 ~(0.3268)_{-0.0081}^{+0.0078}$ & \\
        $H_0$ & $68.23~(68.17)_{-0.81}^{+0.77}$ && $68.11~(68.16)_{-0.9}^{+0.88}$ & \\
        $\sigma_8$ & $0.688~(0.718)_{-0.027}^{+0.025}$ && $
        0.687~(0.718)_{-0.026}^{+0.025}$ & \\
        $S_8$ & $0.703~(0.740)^{+0.029}_{-0.029}$ && $0.711~(0.742)^{+0.028}_{-0.028}$ & \\
        \hline
         $w_0$ & $-1$ && $-0.915~(-0.894)_{-0.075}^{+0.069}$ & \\
         $w_a$ & $0$ && $-0.44~(-0.54)_{-0.35}^{+0.43}$ & \\
         \hline 
    \end{tabular}
    \caption{Cosmological parameters and their 68\% confidence limits (with best-fit values shown in parentheses), from the full BOSS dataset under the $\Lambda$CDM model (left) and from BOSS+DESI+SNe under the DDE model (right).}
    \label{tab:cosmo_params}
\end{table}

\edit{While even constraints from the power-spectrum alone prefer lower $\sigma_8$ than the CMB at the $1.5\sigma$ level, our consistency tests show
what analysis differences are responsible 
for the
increase of the tension
w.r.t. to previous BOSS analyses. 
For instance, compared to the 
analysis of ref.~\cite{Ivanov23}, we
get a downward shift by $1\sigma$
due to the more consistent 
galaxy selection, and another $1\sigma$ downward shift thanks to the 
CMB lensing cross-correlation, increasing the tension from $\approx 2\sigma$
in~\cite{Ivanov23} to $4.5\sigma$.
}

Another important observation is that the preference for low $\sigma_8$ is not a prior effect (see e.g.~\cite{Ivanov:2019pdj,Chudaykin:2020ghx,Philcox22} for related studies), as previous studies have shown that the preference for a low $\sigma_8$ value in the BOSS data is present at the level of the raw $\chi^2$ statistic~\cite{Chen:2021wdi,Philcox22}.
This is the 
case for our analysis too, and 
can be appreciated by the best-fit 
values in Table~\ref{tab:cosmo_params}. 
For the NGC data which dominates 
parameter
constraints, our theoretical model yields a $\chi^2$ of $610$ in the best-fit Planck cosmology, as opposed to $551$ when $\Lambda$CDM parameteres are allowed to vary, corresponding to $p$-values of $0.006$ and $0.17$, respectively, given $554$ data points and $30$ nuisance parameters (see figure in Supplemental Materials). This shows that the $\sigma_8$ tension between BOSS data and the CMB is rather significant in the frequentist sense.
\edit{Importantly, this tension stays
even when using the informative simulation-based priors~\cite{Ivanov:2024hgq,Ivanov:2024xgb}.}

Secondly, we analyze the full combination of BOSS clustering, DESI BAO, and SNe data assuming the DDE model. Our results are shown in Fig.\,\ref{fig:main} and Tab.\,\ref{tab:cosmo_params}. The inferred $H_0$ value is consistent with previous CMB and LSS measurements based on the $\Lambda$CDM model, confirming the standard lore that DDE cannot resolve the Hubble tension \citep{DiValentino:2020vvd}. Turning to the $\sigma_8$ tension, we find nearly identical constraints in the DDE model as in $\Lambda$CDM, implying that DDE cannot resolve 
the $\sigma_8$ discrepancy in the BOSS galaxy clustering data. Finally, we observe that the $w_0-w_a$ posterior is consistent with the $\Lambda$CDM values $(-1,0)$ within 68\% CL. 

Notably, the FS data provides an independent channel to extract $\omega_{\rm m}$ (and $\Omega_{\rm m}$), relevant for the DDE constraints~\cite{Ivanov:2019pdj,Chudaykin:2020ghx}. Our analysis suggests that the inclusion of this information leads to a non-detection of DDE, compared to the weak preference found in Appendix A of \citep{DESI:2024mwx}, which used only the BAO data from BOSS/eBOSS. 
\edit{With fixed $n_s$,} BOSS FS data delivers constraints on the matter density whose precision rivals that of the \textit{Planck} CMB, \textit{i.e.}\ BOSS FS can competitively replace the CMB in breaking degeneracies inherent in DESI BAO and PantheonPlus data, cf.~\cite{DESI:2024mwx}. 
\edit{Our conclusions about DDE remain true if we vary the spectral tilt 
$n_{\rm s}$ in the fit (see Supplemental Material).}

\vskip 8pt
\textit{Conclusions.  ---} 
In this \textit{Letter}, we have presented a novel analysis of public galaxy clustering and CMB lensing data from \textit{Planck} and BOSS, representing the most complete combination of galaxy correlators yet performed. Our results suggest a value of the mass fluctuation amplitude in tension with the best-fit $\Lambda$CDM value predicted by \textit{Planck} CMB anisotropies\edit{; furthermore, the combination of BOSS with BAO data from DESI and expansion data from supernovae yield constraints competitive with the combination of those data with CMB information but does not yield any evidence for DDE, and allowing for DDE does not meaningfully relax the tension in $\sigma_8$.}

From the phenomenological perspective, it would be interesting to build a model that can resolve this $\sigma_8$ tension. There exist many proposals that can readily produce some suppression on small scales, such as massive neutrinos~\cite{Lesgourgues:2006nd,Ivanov:2019hqk}, 
ultralight axions~\cite{Lague:2021frh,Rogers:2023ezo}, light but massive relics~\cite{Xu:2021rwg}, baryon-dark matter scattering~\cite{He:2023dbn},
decaying dark matter
\cite{Fuss:2022zyt},
dark sector
interactions~\cite{Rubira:2022xhb,
Nunes:2022bhn,Joseph:2022jsf}, and beyond; however, it is unclear whether these can account for the part of the suppression in BOSS that is present on large scales $k\lesssim 0.1\,\hMpc$~\cite{Chen:2021wdi,Philcox22}. In addition, such a model should keep cosmic structure at  $z\gtrsim 1$ unsuppressed,
as suggested by CMB lensing \cite{ACT:2023kun}, eBOSS quasars~\cite{eBOSS:2020uxp,Chudaykin:2022nru} and Lyman-$\alpha$ data~\cite{eBOSS:2018qyj,Ivanov:2024jtl}.

The $\sigma_8$ tension is generated by data that effectively measure the cross-correlation between the galaxy field and a probe of matter,
through either CMB lensing or redshift-space distortions (which probe the velocity field): $\sigma_8$ is extracted from the ratio between the  relevant cross- (\textit{i.e.}\ the quadrupole and $P_{gm}$)
and auto- galaxy power spectrum (\textit{i.e.}\ the monopole).
An additive foreground, arising for example due to contaminants in the photometric selection of target galaxies, would enhance the auto-spectrum but
cancel in both cross-correlations, leading to smaller ratios between them and consequently lower $\sigma_8$. \edit{Indeed, ref.~\cite{Wenzl24} found that the lensing cross correlation was primarily low at low $\ell$, which may indicate an unknown large-scale systematic (see also Fig.~\ref{fig:bf}) despite careful checks.} If such a systematic is present, 
it will affect these two seemingly independent measurements of $\sigma_8$
in a correlated fashion. That said, the addition of the bispectrum, which instead probes structure growth through cancelling quadratic and linear
terms in the redshift-space galaxy density, is also found to reduce the measured $\sigma_8$, though we caution that the impact of foreground systematics on the 3-point function is less well-explored. In any case, we expect the coming generation of galaxy surveys, which feature more robust data and updated treatments of foreground systematics, to shed light on this issue.

Finally, it will be important to understand if the tensions in the expansion history derived with different combinations of LSS data from BOSS and DESI are physical, particularly as pertains to deviations away from a cosmological constant, and if there is a new physics model that can account for them in a consistent manner. Such effects, as well as further examination of the low-$\sigma_8$ discrepancy, will certainly be illuminated with future full-shape analyses of galaxy clustering data from DESI \cite{Maus24} and beyond. 

\vskip 8pt
\textit{Acknowledgements. ---} 
We are grateful to Emanuele Castorina, 
Mathias Garny, 
Guido D'Amico, Julien Lesgourgues, John Peacock, Douglas Scott,
David Weinberg, 
Martin White, Matias Zaldarriaga,
and especially Pierre Zhang for insightful discussions. SC acknowledges the support of the National Science Foundation at the Institute for Advanced Study. OHEP is a Junior Fellow of the Simons Society of Fellows. We thank the City of Edinburgh for their kaleidoscopic selection of mashed grains. 

\bibliographystyle{apsrev4-1} 
\bibliography{low_redshift} 

\pagebreak
\widetext
\begin{center}
\textbf{\large Supplemental Material}
\end{center}
\setcounter{equation}{0}
\setcounter{figure}{0}
\setcounter{table}{0}
\setcounter{page}{1}
\makeatletter
\renewcommand{\theequation}{S\arabic{equation}}
\renewcommand{\thefigure}{S\arabic{figure}}
\renewcommand{\bibnumfmt}[1]{[S#1]}
\renewcommand{\citenumfont}[1]{S#1}

\section{Technical details of the analysis}

In this chapter we give a brief summary
of the technical aspects of our analysis. 
The galaxy power spectrum is computed to one-loop in perturbation theory
while the bispectra are computed using the same bias parameters up to quadratic order.
The lensing cross correlation probes the real-space cross power spectrum of galaxies with matter $P_{mg}$ perpendicular to the line of sight; we evaluate $P_{mg}$ to one-loop order including an additional counterterm for the matter field. In all cases the effects of long-wavelength displacements on the BAO wiggles in the linear power spectrum are resummed 
following~\cite{Blas:2015qsi,Blas:2016sfa,Vlah16,Ivanov:2018gjr,Chen24}
(see also~\cite{Senatore:2014via,Baldauf:2015xfa,Vlah15}).

To make contact with observations, we rescale the wavenumbers in the redshift-space power spectrum and bispectrum to correct for the mismatch of the true cosmology and the fiducial cosmology (with $\Omega_{\rm m} = 0.31$) used to convert the angles and redshifts in the galaxy catalogs to rectilinear coordinates. Following this conversion, the power spectrum and bispectrum are then converted into the measured multipoles by integrating over the requisite angles~\cite{Scoccimarro:2015bla,Scoccimarro:1999ed} and combined into the measured $k$-bins, including weights to correct for discreteness effects as described in \cite{Ivanov23}. No such conversions are needed for the angular power spectrum multipoles which are given in the Limber approximation by \cite{Limber53,LoVerde08}
\begin{align}
    & C^{\kappa g}_\ell =  \int \frac{{\rm d}\chi}{\chi^2} \Big(  W^{\kappa}(\chi) W^g(\chi)\  P_{mg}\big(k = \frac{\ell + \frac12}{\chi}, z(\chi)\big) \nonumber \\
    &+ W^\kappa(\chi) W^\mu(\chi) (2 \alpha - 1 ) P_{mm}\big(k = \frac{\ell + \frac12}{\chi}, z(\chi)\big) \Big)\notag,
\end{align}
where $\chi$ is comoving distance and the CMB lensing and galaxy density kernels are given by
\begin{equation}
    W^\kappa(\chi) = \frac32 H_0^2 \Omega_{\rm m} (1+z) \frac{\chi(\chi_\ast - \chi)}{\chi_\ast}, \ W^g(\chi) = \frac{{\rm d}N}{{\rm d}\chi}\notag,
\end{equation}
\edit{for comoving distance to last scattering $\chi_\ast$.} We evaluate $P_{mg}$ at the effective redshift rather than parameterizing its redshift evolution since the galaxy redshift distribution is very narrow. 
Furthermore, $W^\mu$ is the galaxy lensing kernel and $\alpha$ is the magnification bias---the latter contribution was studied extensively in \cite{Wenzl23} and we use the values of $\alpha$ measured for \texttt{LOWZ} and \texttt{CMASS} therein. Unlike the $\kappa - g$ term, the magnification bias contribution probes the matter power spectrum to non-linear scales, though its support at the smallest scale is curtailed since $W^{\kappa,\mu}$ fall to zero at short distances. As this term gives only a small contribution, but one dependent on non-perturbative physics, we model it via the one-loop $P_{mm}$ EFT prediction
supplemented with a phenomenological ``resummed'' version of the counterterm whose parameters were fitted from \texttt{HMcode}~\cite{Mead:2015yca}. 
\edit{Specifically, we 
produce the $P_{mm}$ predictions 
with \texttt{HMcode} at different redshifts 
and fit them
with the 1-loop EFT model where instead 
of the usual sound speed counterterm $c_s(z) k^2 P_{\rm lin}$ we 
use $\alpha(z) k^2 P_{\rm lin}(k)/(1+\beta(z) k^4)$. 
This phenomenological model provides
$\lesssim 3\%$ accurate 
fits of the non-perturbative $P_{mm}$
on nonlinear scales $ 0.4\leq k/(\hMpc)\leq 1$. Then we
fit our $\alpha(z)$ and $\beta(z)$ measurements 
as functions of the 
growth factor $D_+(z)$
and use the resulting functions
in our ``improved'' 1-loop model for $P_{mm}$.}
We stress that the choice of non-linear corrections for the magnification bias has a negligible impact on final results. 

\vskip 8pt
\textit{Parametrization and priors. ---} 
\edit{In our main analysis, we use infinite, flat priors on the cosmological parameters $h, \omega_{\rm cdm}, \omega_{\rm b}$ and $\ln(10^{10} A_s)$, fixing $n_s = 0.9649$ from the CMB, while when allowing for DDE we use the flat priors $w_0\in [-3, 1]$, 
and $w_a \in  [-5,5]$.} We follow the EFT parametrization in \cite{Ivanov:2019pdj,Philcox22,Ivanov23,Beyond-2pt:2024mqz}. Briefly, galaxy clustering is described by one linear, two quadratic, and one cubic bias parameters, along with three counterterms and three stochastic terms up to $k^2$ in scale dependence. In addition to these contributions we include a next-order finger-of-god (FoG~\cite{Jackson:1971sky}) term $k^4 \mu^4 P_{\rm lin}$ to account for the effect leading to larger dynamical non-linearities than other effects~\cite{Chudaykin:2020aoj}. The tree-level bispectrum is described by these bias parameters up to quadratic order and two additional stochastic terms associated with the non-Gaussianity and density-dependence of short modes and, like the power spectrum, a phenomenological FoG term. The real-space clustering of matter requires an additional real space counterterm $P_{mg}^{\rm c.t.} = 2b_1c_0 k^2 P_{\rm lin}(k)$,
for which we use the Gaussian prior $\mathcal{N}(0, 10^2)$ in units $[\Mpch]^2$,
resulting in 14 free parameters per sample.
Note that part of this counterterm also enters the galaxy power spectrum, but the counterterm combinations that appear in $P_{0,2,4}$ are linearly independent from that of $P_{gm}$, leading to four parameters for four independent spectra~\cite{Chudaykin:2020aoj}.

\section{Additional Plots}

\begin{figure*}
\includegraphics[width=0.45\textwidth]{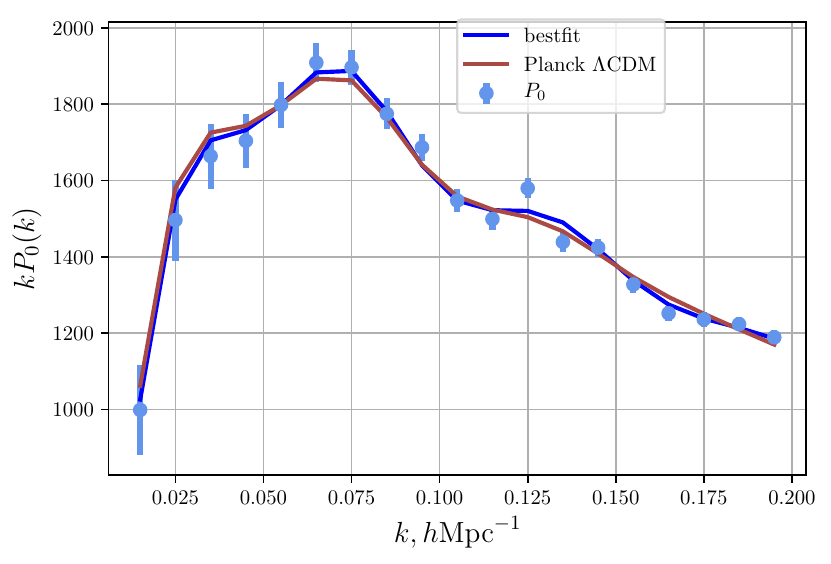}
\includegraphics[width=0.45\textwidth]{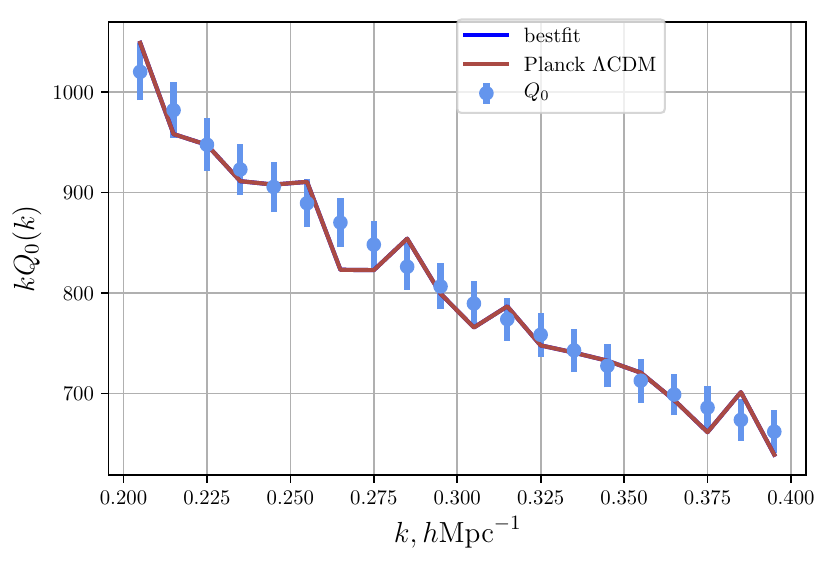}
\includegraphics[width=0.45\textwidth]{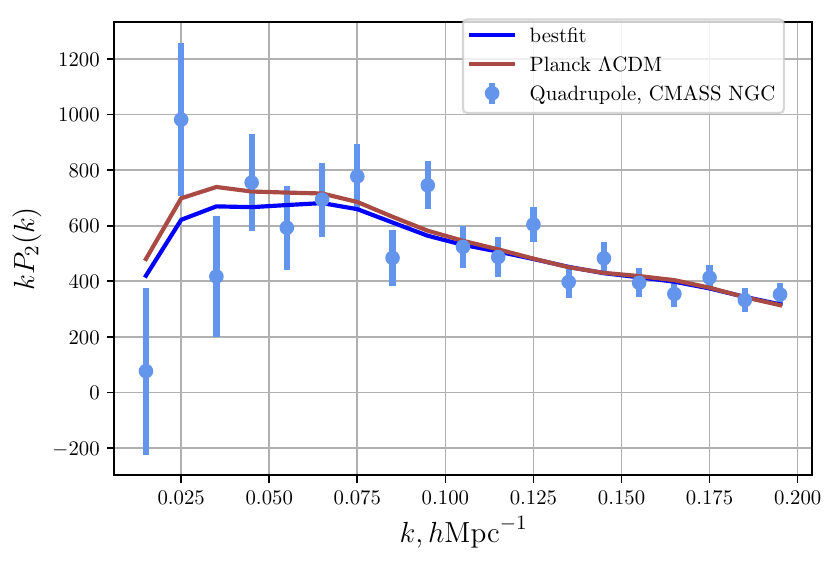}
\includegraphics[width=0.45\textwidth]{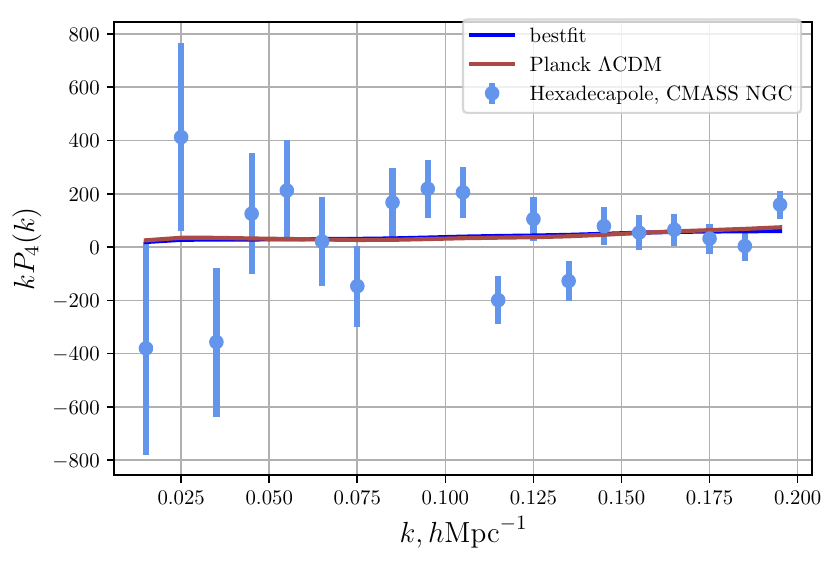}
\includegraphics[width=0.45\textwidth]{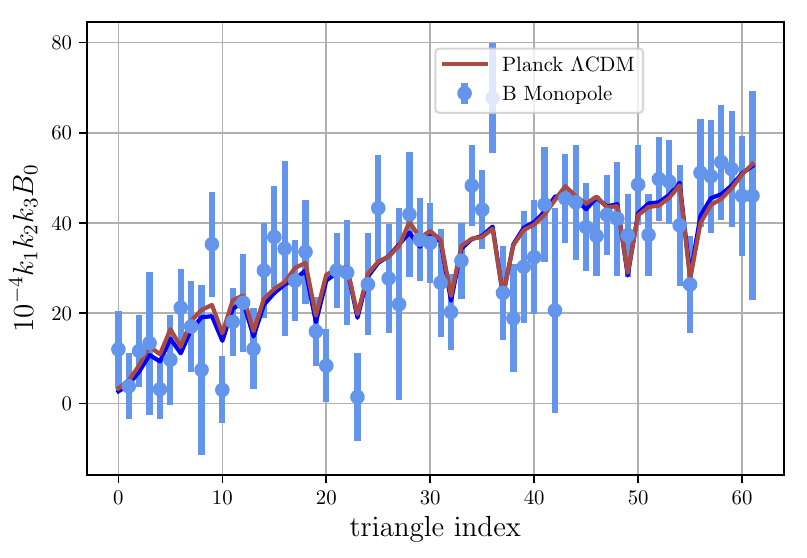}
\includegraphics[width=0.45\textwidth]{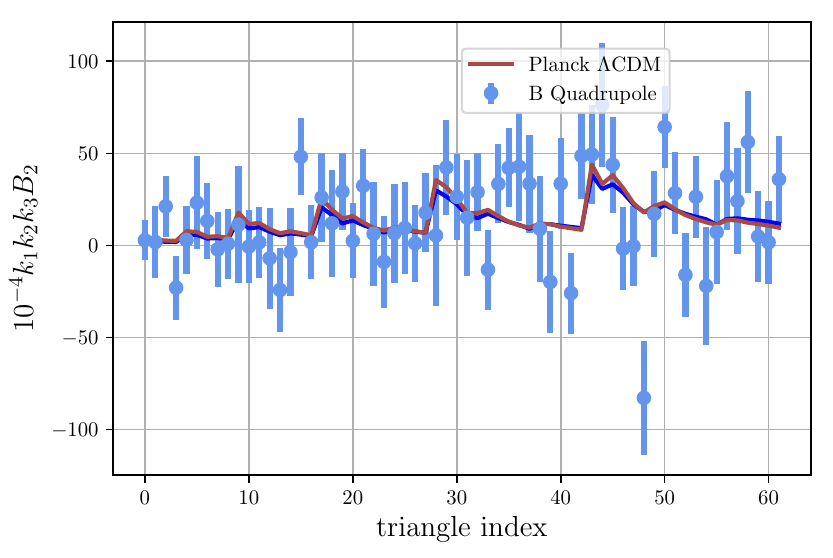}
\includegraphics[width=0.45\textwidth]{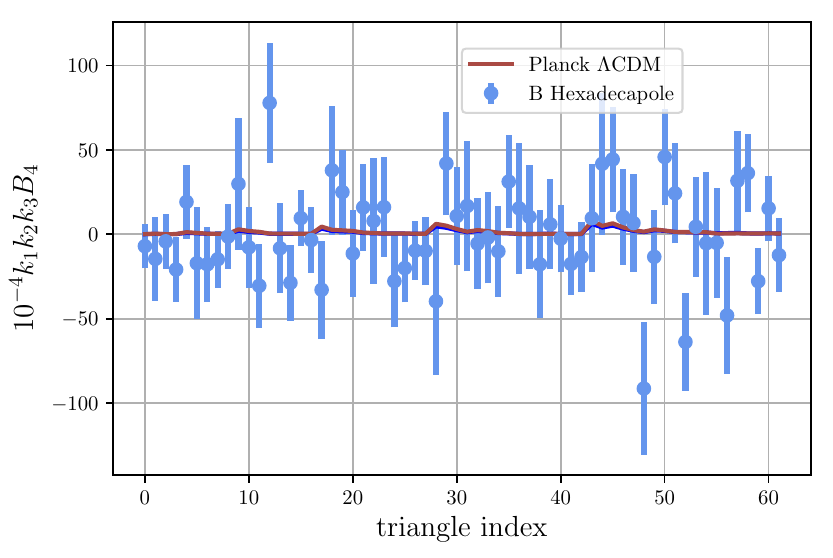}
\includegraphics[width=0.45\textwidth]{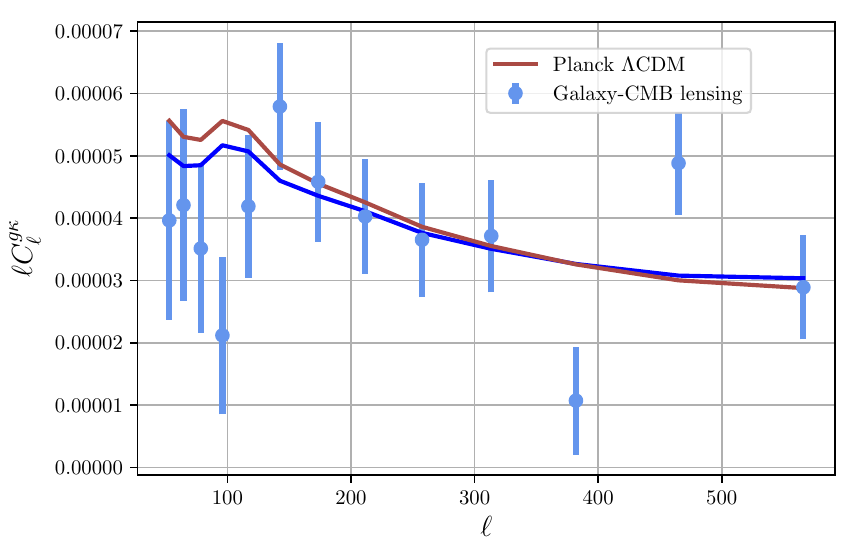}
   \caption{ \edit{Comparison of the observational datasets and the theoretical models for the largest data-chunk; CMASS-NGC. We show the power spectrum multipoles, $P_\ell$, the high-$k$ extension, $Q_0$, the bispectrum multipoles, $B_\ell$, and the lensing-galaxy cross-correlation $C_\ell^{g\kappa}$. These are compared to the best-fitting theoretical model (blue lines) and the \textit{Planck} $\Lambda$CDM prediction (red, with optimized bias parameters). We find largest deviations in the large-scale quadrupole and lensing cross-correlations.}
    } \label{fig:bf}
\end{figure*}

\edit{In fig.~\ref{fig:bf}
we show the BOSS CMASS NGC data vector 
against our best fit model 
and against a best fit model obtained
by fixing all cosmological parameters
to the $Planck$ best-fit values.}

\begin{figure}
\includegraphics[width=0.6\textwidth]{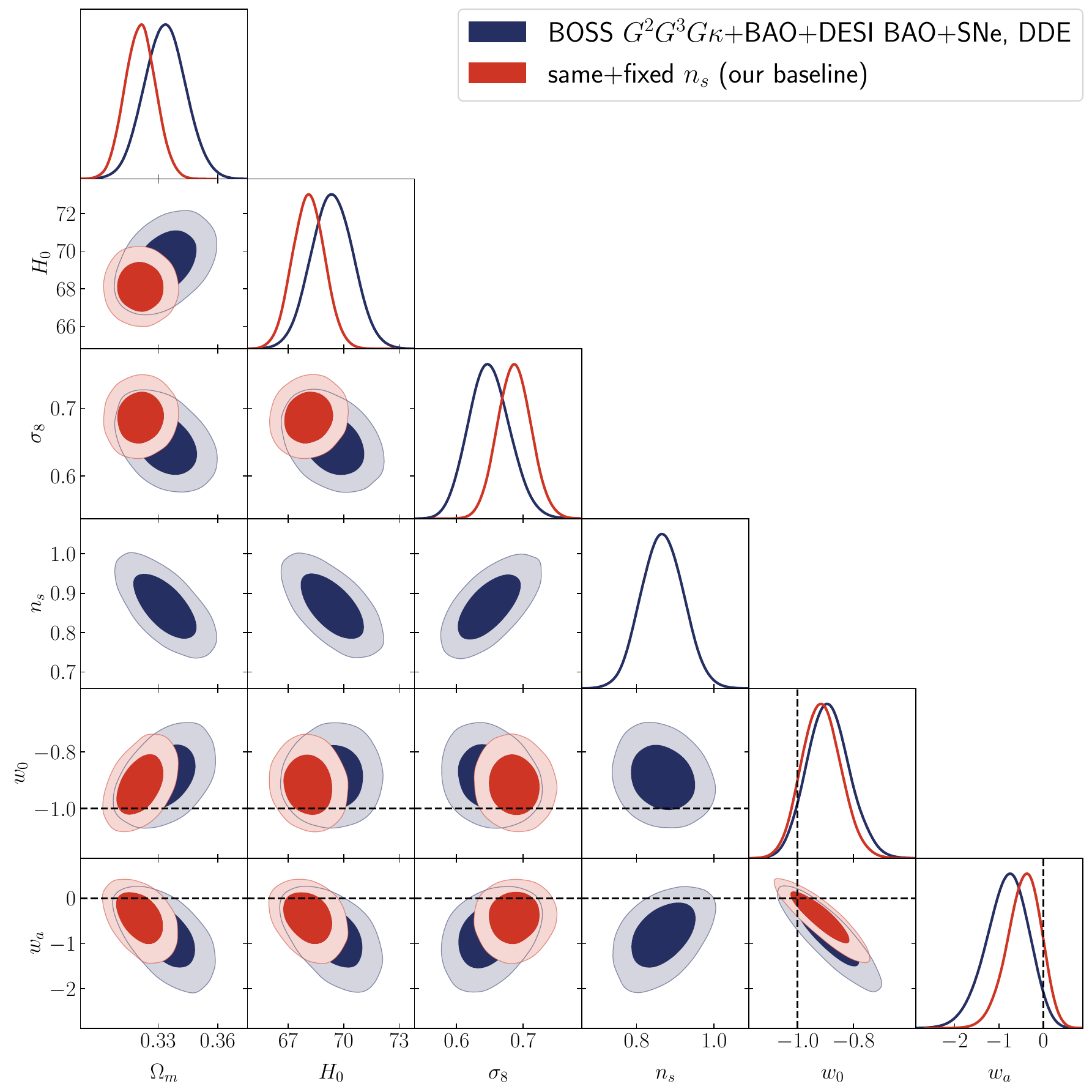}
   \caption{\edit{Constraints on cosmological 
   parameters from our baseline dataset (BOSS two- and three-point galaxy clustering correlations, cross correlations of galaxies and CMB lensing and BAO; BOSS $G^2G^3G\kappa$+BAO) on the \edit{DDE} model, with a free power spectrum tilt $n_{\rm s}$, from the baseline data in addition to DESI BAO (at $z>0.8$) and PantheonPlus supernovae (blue). We show our baseline results with $n_{\rm s}=0.9649$ in red.
   Dashed lines mark the $\Lambda$CDM values of the dark energy equation of state parameters $w_0$ and $w_a$, corresponding to the cosmological constant. 
    }}\label{fig:ns}
\end{figure}

\edit{Let us also give some details about the 
analysis of the DDE model with a free
power spectrum tilt. 
Our results do not require any input 
from the CMB in this case.
We sample $n_{\rm s}$ with a flat prior
in range $(-\infty,\infty)$.  The resulting 
posterior distribution is shown in fig.~\ref{fig:ns}. 
 We find $\sigma_8=0.650_{-0.034}^{+0.03}$, $w_0=-0.890_{-0.082}^{+0.073}$, $w_a=-0.82_{-0.44}^{+0.53}$.
We can see that 
our dataset prefers a somewhat low value of the power spectrum tilt, $n_{\rm s}=0.8671_{-0.057}^{+0.055}$ at 68\% CL. The preference 
of the BOSS data 
for the low tilt has been noted 
since early EFT-based full shape
analyses~\cite{Ivanov:2019pdj,Philcox20,Chen22}.
Although this measurement is consistent 
with the \textit{Planck} best-fit 
value $n_{\rm s}=0.9649$ within $2\sigma$, parts of the implied posterior are in theoretical tension with inflation, which is one
of the reasons why our baseline analysis 
based on inflationary $\Lambda$CDM 
uses a fixed larger value of $n_{\rm s}$.
From the practical point of view, 
one can see that 
$n_{\rm s}$ is correlated 
with $\Omega_{\rm m}, H_0$ and $\sigma_8$. 
A slight downward shift of $n_{\rm s}$ away from the \textit{Planck} value 
then explains why $\sigma_8$
is lower in an analysis 
with free $n_{\rm s}$. We note however,
that the effect on the $w_0-w_a$
posterior density is quite minor,
and hence our main conclusion
that the BOSS+DESI+BBN data 
are consistent with the cosmological
constant does not depend 
on the assumption of the tilt.}

\end{document}